\documentstyle[12pt]{article}

\begin{document}
\hfill {DFTUZ/92/7}

\hfill {May, 1992}
\vspace{1in}
\begin{center}
{\bf REFLECTION POSITIVE FORMULATION OF CHIRAL GAUGE THEORIES
ON A LATTICE} \\
\vspace{0.8in}
{Sergei V. Zenkin}
\footnote{\normalsize E-mail: zenkin@inr.msk.su} \\
\vspace{0.3in}
{Institute for Nuclear Research of Russian Academy of
Sciences, \\
60th October Anniversary Prospect 7a, 117312 Moscow,
Russia}\\
\vspace{1in}
{\bf Abstract} \\
\end{center}
\vspace{0.1in}

Gauge invariant chiral theories satisfying the reflection positivity is
constructed on a lattice. This requires the introduction of "half gauge
fields" defined some time ago by Brydges, Fr\"{o}hlich, and Seiler
\cite{BFS}. A two-dimensional model is considered in some detail.

\newpage

\vspace{0.4in}
{\bf 1. Introduction}
\vspace{0.2in}

Reflection positivity, together with gauge invariance, plays a
fundamental role in constructing continuum quantum gauge theories from
lattice gauge theories (see, e.g., \cite{Seiler} and references therein).
It allows one to construct the Hilbert space of states with positive
definite metric and ensures the canonical quantum mechanical
interpretation of the theories. This guarantees their unitarity, that
is particularly important for gauge theories.

Vector lattice gauge theories like QCD naturally are reflection
positive and gauge invariant, while for chiral ones it is a long
standing problem to satisfy these properties. The only exception is
mirror fermion model \cite{Montvay}, however, it is actually chiral
only in the broken phase of the Higgs sector. It is worth noting that
it is the explicit gauge invariance of this model that makes it
possible to prove its reflection positivity. Models with the
Wilson-Yukawa couplings (see review \cite{Smit17} and references
therein) appear not to be reflection positive \cite{Smit20} and unitary
\cite{Zen1}. Besides, these models do not lead to chiral interacting
continuum theories \cite{GPS}.

In this paper we propose a new formulation of chiral gauge theories on
a lattice which is both reflection positive and gauge invariant.

\vspace{0.4in}
{\bf 2. Basic notations and definitions}
\vspace{0.2in}

Let us first give basic notations and definitions which largely
coincide with those of ref. \cite{BFS}. We consider a hypercubic
D-dimensional (D is even) lattice $\Lambda$ with sites numbered by $n =
(n_{0}, \ldots, n_{D-1})$, $-N/2+1 \le n_{\mu} \le N/2$, $N$ is odd,
with lattice spacing $a$. Let $\hat{\mu}$ be the unit vector along a
lattice link in the positive $\mu$-direction. We shall define a theory
on a torus $T_{D}$ which is obtained by the addition of links
connecting each pair of sites with $n_{\mu} = N/2$ and $n_{\mu} =
-N/2+1$. Thus, our lattice has $N^{D}$ sites and $(D N)^{D}$ links.

Let $G = G^{L} \times G^{R}$ be the gauge group which we consider to be
unitary. Dynamical variables of the theory are the fermion
(Grassmannian) fields $\psi_{n}$, $\overline{\psi}_{n}$, defined on
lattice sites, and the gauge variables $U^{L,R}_{n, n + \hat{\mu}} \in
G^{L,R}$, $U_{n, n - \hat{\mu}} = U^{\dag}_{n -\hat{\mu}, n}$ defined
on lattice links. We shall use the representation for $U^{L,R}$
introduced in ref. \cite{BFS} in terms of "half gauge" variables
$W^{L,R}_{(n, \pm \hat{\mu})} \in G^{L,R}$ associated with each pair
$(n, \pm \hat{\mu})$:
\begin{equation}
U^{L,R}_{n, n + \hat{\mu}} = W^{L,R}_{(n, \hat{\mu})} \;
{W^{L,R}_{(n + \hat{\mu}, -\hat{\mu})}}^{\dag}.
\end{equation}
($W_{(n, \pm \hat{\mu})}$ is $w^{\dag}_{n, <n, n \pm \hat{\mu}>}$ of
ref. \cite{BFS}.)

Gauge transformations are defined as follows:
\begin{eqnarray}
\psi_{n} \rightarrow (h^{L}_{n} P_{L} + h^{R}_{n} P_{R}) \psi_{n},
& & \overline{\psi}_{n} \rightarrow \overline{\psi}_{n}
({h^{L}_{n}}^{\dag} P_{R} + {h^{R}_{n}}^{\dag} P_{L}),  \\ U^{L,R}_{m,
n} \rightarrow h^{L,R}_{m} \; U^{L,R}_{m, n} \; {h^{L,R}_{n}}^{\dag}, &
& W^{L,R}_{(n, \pm \hat{\mu})} \rightarrow h^{L,R}_{n} \; W^{L,R}_{(n,
\pm \hat{\mu})},
\end{eqnarray}
where $h^{L,R}_{n} \in G^{L,R}$, $P_{L,R} = (1 \pm \gamma_{D+1})/2$,
$\gamma_{D+1} = - i^{D/2} \gamma_{0} \cdots \gamma_{D-1}$ (we use
Hermitean $\gamma$-matrices with $[\gamma_{\mu}, \gamma_{\nu}]_{+} = 2
\delta_{\mu \nu}$).

Let $\Lambda_{\pm}$ denote the equal parts of the lattice with $n_{0} >
0$ and $n_{0} < 0$, respectively, and let $r$ be such a reflection,
that $r \Lambda_{\pm} = \Lambda_{\mp}$. So, the reflection do not
change $n_{\mu}$ and $\hat{\mu}$ for $\mu \neq 0$, while $r n_{0} = -
n_{0} + 1$, $r \hat{0} = - \hat{0}$.

Given reflection $r$, an antilinear operator $\theta$ is defined such
as
\begin{equation}
\theta [\overline{\psi}_{m} \: \Gamma \: U_{m, n} \cdots W_{(n, \pm
\hat{\mu})} \cdots \psi_{n}] = \overline{\psi}_{r n} \gamma_{0} \cdots
W^{\dag}_{r (n, \pm \hat{\mu})} \cdots U_{r n, r m} \: \Gamma^{\dag} \:
\gamma_{0} \psi_{r m},
\end{equation}
where $\Gamma$ is a matrix.

A theory with an action $A[\psi, \overline{\psi}, W]$ is called
reflection positive if for each functional $F[\psi, \overline{\psi},
W]$ defined on $\Lambda_{+}$ one has
\begin{equation}
\int \prod_{n \in \Lambda} d \psi_{n} d \overline{\psi}_{n} \prod_{n
\in \Lambda, \mu} d W_{(n, \hat{\mu})} d W_{(n, -\hat{\mu})} \; F
\theta [F] \; e^{-\displaystyle A} \geq 0,
\end{equation}
where $d W_{(n, \pm \hat{\mu})}$ denotes the Haar measure.

The sufficient condition for a theory with an action $A$ to be
reflection positive is existing such functionals $B[\psi,
\overline{\psi}, W]$ and $C_{i}[\psi, \overline{\psi}, W]$ defined on
$\Lambda_{+}$ that $A$ can be represented in the form \cite{BFS}
\begin{equation}
- A = B + \theta [B] + \sum_{i} C_{i} \theta [C_{i}].
\end{equation}

\vspace{0.4in}
{\bf 3. Constructing the theory}
\vspace{0.2in}

We proceed from the lattice action for free massless fermions
of the form \cite{Zen2}
\begin{equation}
A = a^{D} \sum_{n \in \Lambda, \mu} \overline{\psi}_{n} \biggl[
\gamma_{\mu} \frac{1}{2a} (\psi_{n +
\hat{\mu}} - \psi_{n - \hat{\mu}})
- T \frac{1}{2a} (\psi_{n + \hat{\mu}} + \psi_{n -
\hat{\mu}} - 2 \psi_{n})\biggr],
\end{equation}
where $\psi_{(\ldots, N/2 + 1, \ldots)} = - \psi_{(\ldots, -N/2 + 1,
\ldots)}$, $\psi_{(\ldots, -N/2, \ldots)} = - \psi_{(\ldots, N/2,
\ldots)}$; $T$ is an unitary matrix under constraint $T^{\dag} =
\gamma_{0} T \gamma_{0}$. This is the general form of the lattice
fermion action which is determined by the finite dimension
approximation of functional integrals for canonical Hamiltonian
(Grassmannian) dynamics and satisfies condition (6) of reflection
positivity; at $T = 1$ it is exactly the Wilson fermion action.

It is not invariant under the global transformation of the form (2).
Therefore, we seek the gauge action in the form
\begin{eqnarray}
fA = a^{D} \sum_{n \in \Lambda, \mu} \overline{\psi}_{n} \biggl[
\gamma_{\mu}\frac{1}{2a}&\biggl(& (P_{L}U^{L}_{n, n+\hat{\mu}} +
P_{R}U^{R}_{n, n+\hat{\mu}}) \psi_{n +\hat{\mu}} -
(P_{L}U^{L}_{n, n-\hat{\mu}} +
P_{R}U^{R}_{n, n-\hat{\mu}}) \psi_{n - \hat{\mu}}\biggr) \nonumber \\ -
\frac{1}{2a}&\biggl(& (P_{L}X^{L}_{n, n+\hat{\mu}} + P_{R}X^{R}_{n,
n+\hat{\mu}})\psi_{n + \hat{\mu}} + (P_{L}X^{L}_{n, n-\hat{\mu}} +
P_{R}X^{R}_{n, n-\hat{\mu}})
\psi_{n - \hat{\mu}} \nonumber \\
                        &       &- 2 (P_{L}Y^{L}_{(n,
\hat{\mu})} +
P_{R}Y^{R}_{(n, \hat{\mu})}) \psi_{n}\biggr)
\biggr],
\end{eqnarray}
where $X$ and $Y$ are some functions of $W$. Let us now require the
action (8) to be invariant under rotations of the lattice by $\pi/2$,
and to satisfy the following conditions:

(i) condition (6) of reflection positivity;

(ii) gauge invariance;

(iii) for vector group $G^{L} = G^{R}$ at $T = 1$ it takes the familiar
Wilson form;

(iv) in the ungauged limit $W =1$ it takes the form of eq. (7).\\
Then from (i) and (iv) we find
\begin{eqnarray}
X^{L}_{n, n + \hat{\mu}} = W^{R}_{(n, \hat{\mu})}\, T \,{W^{L}_{(n
+ \hat{\mu},
-\hat{\mu})}}^{\dag},
& &
X^{L}_{n, n - \hat{\mu}} = W^{R}_{(n, -\hat{\mu})}\, T \,{W^{L}_{(n
- \hat{\mu},
\hat{\mu})}}^{\dag},
\nonumber \\
X^{R}_{n, n + \hat{\mu}} = W^{L}_{(n, \hat{\mu})}\, T \,{W^{R}_{(n
+ \hat{\mu},
-\hat{\mu})}}^{\dag},
& &
X^{R}_{n, n - \hat{\mu}} = W^{L}_{(n, -\hat{\mu})}\, T \,{W^{R}_{(n
- \hat{\mu},
\hat{\mu})}}^{\dag},
\end{eqnarray}
while from (ii) -- (iv) one has
\begin{eqnarray}
&Y^{L}_{(n,\hat{\mu})}& = \frac{1}{2}\biggl( W^{R}_{(n, \hat{\mu})}\,T
\,{W^{L}_{(n, \hat{\mu})}}^{\dag} + W^{R}_{(n, - \hat{\mu})}\, T
\,{W^{L}_{(n, -\hat{\mu})}}^{\dag}\biggr), \nonumber \\
&Y^{R}_{(n,\hat{\mu})}& = \frac{1}{2}\biggl( W^{L}_{(n, \hat{\mu})}\,T
\,{W^{R}_{(n, \hat{\mu})}}^{\dag} + W^{L}_{(n, - \hat{\mu})}\, T \,
{W^{R}_{(n,
-\hat{\mu})}}^{\dag}\biggr).
\end{eqnarray}
So these requirements determine the action uniquely. The important
point, which we come to by product, is that the Feynman rules for
action (8) at $a \rightarrow 0$ coincides with ones for the
corresponding continuum theory. Besides, in a pure chiral case, say
$G^{R} = 1$, the action has the Golterman-Petcher symmetry
\cite{GP}, i.e. is invariant under the global transformations
$\psi_{n} \rightarrow \psi_{n} + P_{R} \epsilon$, $\overline{\psi}_{n}
\rightarrow \overline{\psi}_{n} + \overline{\epsilon} P_{L}$, that
guarantees the decoupling of the right-handed fermions in the continuum
limit.

The full action contains pure gauge part which we require to satisfy
the points (i) -- (iii). Then it is also determined in fact uniquely
and has the form
\begin{equation}
A_{gauge} = \sum_{Plaquettes \in \Lambda} \biggl(\beta_{L} \,
A_{P}[U^{L}] +
\beta_{R} \, A_{P}[U^{R}]\biggr) + \sum_{n \in \Lambda, \mu} \zeta \,
Tr \biggl(Y^{L}_{(n,\hat{\mu})} {Y^{L}_{(n,\hat{\mu})}}^{\dag} - 1
\biggr),
\end{equation}
where $A_{P}[U]$ are the Wilson plaquette action.

One can rewrite actions (8) and (11), and the measure in the functional
integrals in terms of $U^{L,R}_{n, n \pm \hat{\mu}}$ and $W^{L,R}_{(n,
\hat{\mu})}$. Then our theory is determined by functional integrals of
the form
\begin{eqnarray}
Z^{-1} \int \prod_{n \in \Lambda} d \psi_{n} d \overline{\psi}_{n}
\prod_{n \in \Lambda, \mu} d U^{L}_{n, n + \hat{\mu}} d U^{R}_{n, n +
\hat{\mu}} d W^{L}_{(n, \hat{\mu})} d W^{R}_{(n, \hat{\mu})} \;
F[\psi, \overline{\psi}, U^{L}, U^{R}] \nonumber \\
e^{-\displaystyle A[\psi, \overline{\psi}, U^{L}, U^{R}, W^{L}, W^{R}]},
\end{eqnarray}
where $Z$ is partition function of the theory, $F$ is a gauge invariant
functional of the dynamical variables, and $A$ is three parametric
action which is sum of (8) and (11). Thus, the fundamental difference
of chiral gauge theories from vector ones is untrivial integration over
variables $W^{L,R}_{(n, \hat{\mu})}$ which in this case cannot be
cancel out from the action by any way, including a gauge fixing.

\vspace{0.4in}
{\bf 4. Two-dimensional model}
\vspace{0.2in}

To demonstrate how such a formulation works, we consider in more detail
exactly solvable model which is a generalization of the chiral
Schwinger model \cite{JR}. Let $D =2$, $G = U(1)$, $g$ be the gauge
coupling; fermions carry an isotopic index, and left-handed and
right-handed fermions have charges $g Q_{L}$ and $g Q_{R}$,
respectively, where $Q_{L,R}$ some diagonal matrices; $T =1$. Then
$U^{L,R}_{n, n + \hat{\mu}} = (U_{n, n + \hat{\mu}})^{Q_{L,R}}$, $U_{n, n
+ \hat{\mu}} = \exp[iga A_{\mu}(n + \frac{1}{2} \hat{\mu})]$, where
$A_{\mu}$ is the gauge field, and $W^{L,R}_{(n, \hat{\mu})} =
(W_{(n, \hat{\mu})})^{Q_{L,R}}$. For variable $W$ there is no a
representation in terms of a local field which transforms as an
irreducible representation of the group of rotation of the Euclidean
space. However, to allow for a perturbative consideration we introduce
variables $z_{\hat{\mu}}(n)$ so that $W_{(n, \hat{\mu})} = \exp[iga
z_{\hat{\mu}}(n)]$. Then the gauge transformations looks as
\begin{equation}
A_{\mu}(n + \frac{1}{2} \hat{\mu}) \rightarrow
A_{\mu}(n + \frac{1}{2} \hat{\mu}) + \frac{1}{a}(\alpha_{n + \hat{\mu}}
- \alpha_{n} ), \; \; \; \; z_{\hat{\mu}}(n) \rightarrow
z_{\hat{\mu}}(n) - \frac{1}{a} \alpha_{n},
\end{equation}
where $h_{n} = \exp(-ig \alpha_{n})$.

Let us consider the continuum limit of the effective actions
\begin{equation}
W[U] = - \ln \int \prod_{n \in \Lambda} d \psi_{n} d
\overline{\psi}_{n} \prod_{n \in \Lambda, \mu} d W_{(n, \hat{\mu})} \;
e^{-\displaystyle A[\psi, \overline{\psi}, U, W]}.
\end{equation}
We divide the calculation into two steps. First we calculate the
effective action
\begin{equation}
W[U, W] = - \ln \int \prod_{n \in \Lambda} d \psi_{n} d
\overline{\psi}_{n} \; e^{-\displaystyle A[\psi, \overline{\psi}, U,
W]}.
\end{equation}
While a finite $a$ expression for $W[U, W]$ is very cumbersome and
hardly tractable, its continuum limit in terms of $A$ and $z$ can be
found in the closed form. The crucial simplification comes from using
the Reisz theorems \cite{R}. So, for sufficient smooth variables $A$
and $z$ we find
\begin{eqnarray}
W[A, z]&=&\frac{1}{2} \int \frac{d^{2} q}{(2 \pi)^2}
\sum_{\mu, \nu} \biggl[ A_{\mu}(q) \; \biggl( \delta_{\mu \nu} q^{2} -
q_{\mu} q_{\nu} + \frac{1}{2} Tr(Q^{2}_{L}+Q^{2}_{R}) \; \;
\Pi^{AA}_{\mu \nu}(q) \biggr) \; A_{\nu}(q) \nonumber \\
       & &\mbox{} + \frac{1}{2} Tr(Q_{L}-Q_{R})^{2} \; \;
\biggl(
\bar{A}_{\mu}(q)\; (- \zeta g^{2} \; \delta_{\mu \nu} +
\Pi^{\bar{A} \bar{A}}_{\mu \nu})\; \bar{A}_{\nu}(q) \nonumber \\
       & &\mbox{}
+ \bar{A}_{\mu}(q) \;\Pi^{\bar{A} z}_{\mu \nu}\;
z_{\hat{\nu}}(q)
+ z_{\hat{\mu}}(q) \;\Pi^{z \bar{A}}_{\mu \nu}\; \bar{A}_{\nu}(q)
+ z_{\hat{\mu}}(q) \;\Pi^{z z}_{\mu \nu} \;z_{\hat{\nu}}(q)
\biggr) \nonumber \\
       & &\mbox{}+ \frac{i}{2} Tr(Q^{2}_{L}-Q^{2}_{R}) \;\;
\bar{A}_{\mu}(q)\; K^{\bar{A} \bar{A}}_{\mu \nu}(q) \;\bar{A}_{\nu}(q)
\biggr].
\end{eqnarray}
Here $\bar{A}_{\mu}(q)$ is $a \rightarrow 0$ limit of the gauge
invariant combination ${A}_{\mu}(q) + 2i \sin (\frac{1}{2}q_{\mu}a) \;
z_{\hat{\mu}}(q)$,
\begin{equation}
\Pi^{AA}_{\mu \nu}(q) = \frac{g^{2}}{\pi} (\delta_{\mu \nu} -
\frac{q_{\mu} q_{\nu}}{q^{2}}),
\; \; \; \;
K^{\bar{A} \bar{A}}_{\mu \nu}(q) = \frac{g^{2}}{2 \pi}
\frac{1}{q^{2}} (\epsilon_{\mu
\alpha} q_{\alpha} q_{\nu} + q_{\mu} \epsilon_{\nu \alpha} q_{\alpha}),
\end{equation}
and $\Pi^{\bar{A} \bar{A}}_{\mu \nu}$, $\Pi^{\bar{A} z}_{\mu \nu} =
\Pi^{z \bar{A}}_{\mu \nu}$, and $\Pi^{z z}_{\mu \nu}$ are some
symmetrical matrices independent of $q$. From the Ward identities it
follows $\sum_{\mu} \Pi^{z \cdot}_{\mu \nu} = \sum_{\nu} \Pi^{\cdot
z}_{\mu \nu} = 0$, so, effective action (16) is explicitly gauge
invariant. The following relations are also satisfied:
\begin{equation}
\Pi_{0 0} = \Pi_{1 1}, \;\;\;\;
\Pi^{\bar{A} z}_{\mu \nu} = - \frac{1}{2} \Pi^{z z}_{\mu \nu}, \;\;\;\;
\Pi^{\bar{A} z}_{0 1} = - 2 \Pi^{\bar{A} \bar{A}}_{0 1}.
\end{equation}
Numerically the elements of these matrices depend on infrared
regularization used and have not a deep meaning; we started with the
finite lattice which ensures such a regularization, then $\Pi^{\bar{A}
\bar{A}}_{0 0} = 1.959(7) g^{2}/(2 \pi)$, $\Pi^{\bar{A} \bar{A}}_{0 1}
= - 0.361(6) g^{2}/(2 \pi)$.  Obviously expression (16) has no
continuum rotational symmetry.

Now we can represent the continuum limit of effective action (14) as
\begin{equation}
W[A] = - \ln \int \prod_{q, \mu} d z_{\hat{\mu}}(q) \;
e^{-\displaystyle W[A, z]}.
\end{equation}
As $\det \Pi^{z z}_{\mu \nu} = 0$, we have to fix a gauge. Let us
choose $z_{\hat{0}} = 0$. Then, performing the Gaussian integration
with taking into account (18) we find finally
\begin{eqnarray}
W[A]&=&\frac{1}{2} \int \frac{d^{2} q}{(2 \pi)^2} A_{\mu}(q) \; \biggl[
\delta_{\mu \nu} q^{2} - q_{\mu} q_{\nu} + \frac{1}{2}
Tr(Q^{2}_{L}+Q^{2}_{R})  \;
\Pi^{AA}_{\mu \nu}(q) \nonumber \\
    & &\mbox{}+ \frac{1}{2} Tr(Q_{L}-Q_{R})^{2} \;
(- \zeta g^{2} + \Pi^{\bar{A} \bar{A}} )\; \delta_{\mu \nu}
+ \frac{i}{2} Tr(Q^{2}_{L}-Q^{2}_{R}) \;
K^{\bar{A} \bar{A}}_{\mu \nu}(q) \biggr] \;
A_{\nu}(q),
\end{eqnarray}
where $\Pi^{\bar{A} \bar{A}} = \Pi^{\bar{A} \bar{A}}_{0 0} +
\Pi^{\bar{A} \bar{A}}_{0 1}$.  This is exactly what one would expect.
This expression is rotational invariant and reproduces all well-known
features of the vector Schwinger model $(Q_{L} = Q_{R})$ and purely
chiral one $(Q_{R} = 0)$, including both mechanisms of gauge boson mass
generation and two-dimensional anomaly $K^{\bar{A} \bar{A}}(q)$
\cite{JR}.

Let us note that despite the second term in the action (11) the gauge
boson can acquire a mass only in the presence of the fermions, i.e.
when $\Pi \ne 0$. Then the mass can be tuned by parameter $\zeta$.
Otherwise, it remains massless. Indeed, integrating over
$z_{\hat{\mu}}$ in (19) at $\Pi = K = 0$ and taking into account the
trace of $z$ in $\bar{A}$ we find that the counterpart of $\zeta$ is
canceled out from $W[A]$.

\vspace{0.4in}
{\bf 5. Conclusion}
\vspace{0.2in}

Both reflection positivity and gauge invariance of our formulation of
chiral gauge theories on a lattice are due to the use of "half gauge"
variables $W_{(n, \pm \hat{\mu})}$ in terms of which gauge variables
$U$ are represented. The theory can be rewritten in terms of $U$ and,
e.g., $W_{(n, \hat{\mu})}$. In the vector limit $G^{L} = G^{R}$ the
latter vanish in the action, so that integration over them becomes
trivial at arbitrary $a$. In the general case of chiral theories
integration over $W$ is not factorized even at $a \rightarrow 0$, as
our two-dimensional example demonstrates. Moreover, this integration is
found to be necessary for restoring the rotation invariance of the
theory in the continuum limit.

Certainly, considered two-dimension model is only the first test of
this formulation and many questions are still to be answered for
establishing its status. The principal ones concern phase structure and
the continuum limit of four-dimensional theories. However, the fact
that this formulation satisfies all basic requirements as invariance
under the lattice translations and rotations, reflection positivity,
and gauge invariance, is good reason to hope that it ensures the
correct nonperturbative definition of the chiral gauge theories,
including asymptotically free ones and realistic models.

\vspace{0.4in}
{\bf Acknowledgments}
\vspace{0.2in}

I am grateful to M. I. Polikarpov and S. V. Shabanov for useful
discussions at the early stage of the work, and to J. L. Alonso,
M. Asorey, J. L. Cort\'{e}s, and J. G. Esteve for interesting discussion
at the final stage of the work.

It is pleasure to thank J. L. Alonso and Departamento de F\'{\i}sica
Te\'{o}rica de la Facultad de Ciencias de la Universidad de Zaragoza,
where this work has been completed, for their warm hospitality.

\end{document}